\documentclass{ifacconf}

\usepackage{amssymb}
\usepackage{amsmath,amsfonts}
\usepackage{bm}
\usepackage{listings}
\usepackage{color}
\usepackage{mathtools}
\usepackage[version=4]{mhchem}
\usepackage{comment}
\usepackage{makecell}
\usepackage{siunitx}

\theoremstyle{plain}
\newtheorem{theorem}{Theorem}

\newtheorem{proposition}{Proposition}

\theoremstyle{definition}

\newtheorem{assumption}{Assumption}

\newenvironment{proof}[1][Proof]{%
  \par\noindent\textbf{#1.}\ }{%
  \hfill$\square$\par}

\usepackage{graphicx}      
\usepackage{natbib}        
\begin{document}
\begin{frontmatter}

\title{Model Reduction of Multicellular Communication Systems via Singular Perturbation: Sender–Receiver Systems\thanksref{footnoteinfo}\thanksref{arxivnote}} 

\thanks[footnoteinfo]{This work was supported in part by the JSPS Overseas Research Fellowships, an NSF CAREER Award 2240176, the Army Young Investigator Program Award W911NF2010165 and the Institute of Collaborative Biotechnologies/Army Research Office grants W911NF19D0001, W911NF22F0005, W911NF1920026, and W911NF2320006. This work was also supported in part by a subcontract awarded by the Pacific Northwest National Laboratory for the Secure Biosystems Design Science Focus Area “Persistence Control of Engineered Functions in Complex Soil Microbiomes" sponsored by the U.S. Department of Energy Office of Biological and Environmental Research.}
\thanks[arxivnote]{This work has been submitted to IFAC for possible publication.}

\author[First]{Taishi Kotsuka} 
\author[Second]{Enoch Yeung} 

\address[First]{Department of Mechanical Engineering, University of California at Santa Barbara, Santa Barbara, CA 93106 USA (e-mail: tkotsuka@ucsb.edu).}
\address[Second]{Department of Mechanical Engineering, University of California at Santa Barbara, Santa Barbara, CA 93106 USA (e-mail: eyeung@ucsb.edu).}

\begin{abstract}                
We investigate multicellular sender–receiver systems embedded in hydrogel beads, where diffusible signals mediate interactions among heterogeneous cells. Such systems are modeled by PDE–ODE couplings that combine three-dimensional diffusion with nonlinear intracellular dynamics, making analysis and simulation challenging. We show that the diffusion dynamics converges exponentially to a quasi-steady spatial profile and use singular perturbation theory to reduce the model to a finite-dimensional multi-agent network. A closed-form communication matrix derived from the spherical Green’s function captures the effective sender-receiver coupling. Numerical results show the reduced model closely matches the full dynamics while enabling scalable simulation of large cell populations.
\end{abstract}


\end{frontmatter}

\section{INTRODUCTION}

Multicellular communication systems---where individual cells exchange biochemical information through diffusible signal molecules---play essential roles in natural processes such as quorum sensing, morphogenesis, and cooperative metabolic regulation \cite{Miller2001, BenJacob2004}. These principles have been increasingly adopted in synthetic biology, where engineered cell populations exploit sender--receiver signaling architectures to implement distributed sensing, decision-making, and coordinated gene expression \cite{Basu2005}. Beyond their biological significance, such systems carry substantial engineering potential: they enable programmable spatial computation, modular population-level behaviors, and robust task allocation in heterogeneous cellular communities. Among various implementation platforms, hydrogel-embedded multicellular systems have gained particular attention since hydrogels provide a biocompatible, mechanically stable scaffold with tunable diffusion properties and precise spatial organization of cells \cite{sousa2023hierarchical,yoshida2017comp,KURASHINA20211simul,Jeong2023}. Hydrogel-encapsulated sender--receiver systems therefore hold promise for implantable living diagnostics, smart therapeutic biomaterials, biosensing beads, and distributed bioprocessing units, motivating the need for rigorous modeling and analysis of their communication dynamics.

From a systems and control perspective, these multicellular constructs can be naturally interpreted as heterogeneous multi-agent systems in which agents interact indirectly through a shared physical communication channel. Sender cells act as active signaling agents that release diffusible molecules, while receiver cells respond to the aggregate biochemical influence created by the sender population. This viewpoint enables the application of powerful control-theoretic methodologies such as Lyapunov and input--output analysis \cite{Khalil2002,Sontag2008}, and graph-theoretic approaches to stability, synchronization, and collective behavior \cite{Olfati2004,Mesbahi2010graph}. A mathematically grounded multi-agent viewpoint can thus provide design principles for engineering multicellular systems with predictable spatial responses, robustness to heterogeneity, and tunable communication properties.

Despite these advantages, analyzing diffusion-mediated sender--receiver systems remains challenging because the underlying dynamics consist of PDE--ODE couplings that combine a three-dimensional diffusion field with intracellular nonlinear reactions. While full numerical simulations provide detailed spatiotemporal profiles, they become computationally burdensome for large cell populations, even with optimized coupling schemes \cite{Carraro2016}. Conversely, analytical approaches based on matched asymptotic expansions have successfully reduced these systems to lower-dimensional models \cite{Ridgway2022, Bressloff2024}. However, these reductions are typically limited to steady-state analysis or restrict the dynamic analysis to the limit of large diffusivity (well-mixed approximation), thereby failing to capture transient behaviors where spatial gradients play a dominant role. These limitations underscore the need for principled model reduction techniques that preserve essential biophysical heterogeneity and transient spatial effects while enabling tractable analysis and computation.

In this paper, we develop such a framework for hydrogel-embedded multicellular sender--receiver systems. We first establish that the diffusion dynamics converges exponentially to its quasi-steady spatial profile, which allows the application of singular perturbation theory \cite{Khalil2002} to eliminate the fast PDE dynamics. Building on this property, we derive a reduced-order network model in which the intracellular signaling state satisfies a static relation, where the communication matrix is obtained analytically from the Green's function of the spherical diffusion operator \cite{jackson_classical_1999}. This reduction transforms the original PDE--ODE system into a finite-dimensional heterogeneous multi-agent network amenable to standard control-theoretic analysis. Numerical simulations further demonstrate that the reduced model faithfully approximates the full system, even in scenarios involving thousands of spatially distributed cells. The resulting framework provides a scalable and analytically tractable foundation for understanding and engineering diffusion-mediated multicellular communication systems.

\section{Multi-Agent diffusion systems with heterogeneous cells}
\label{sec:modeling}

\subsection{Physical problem setup}

\begin{figure}
    \centering
    \includegraphics[width=0.99\linewidth]{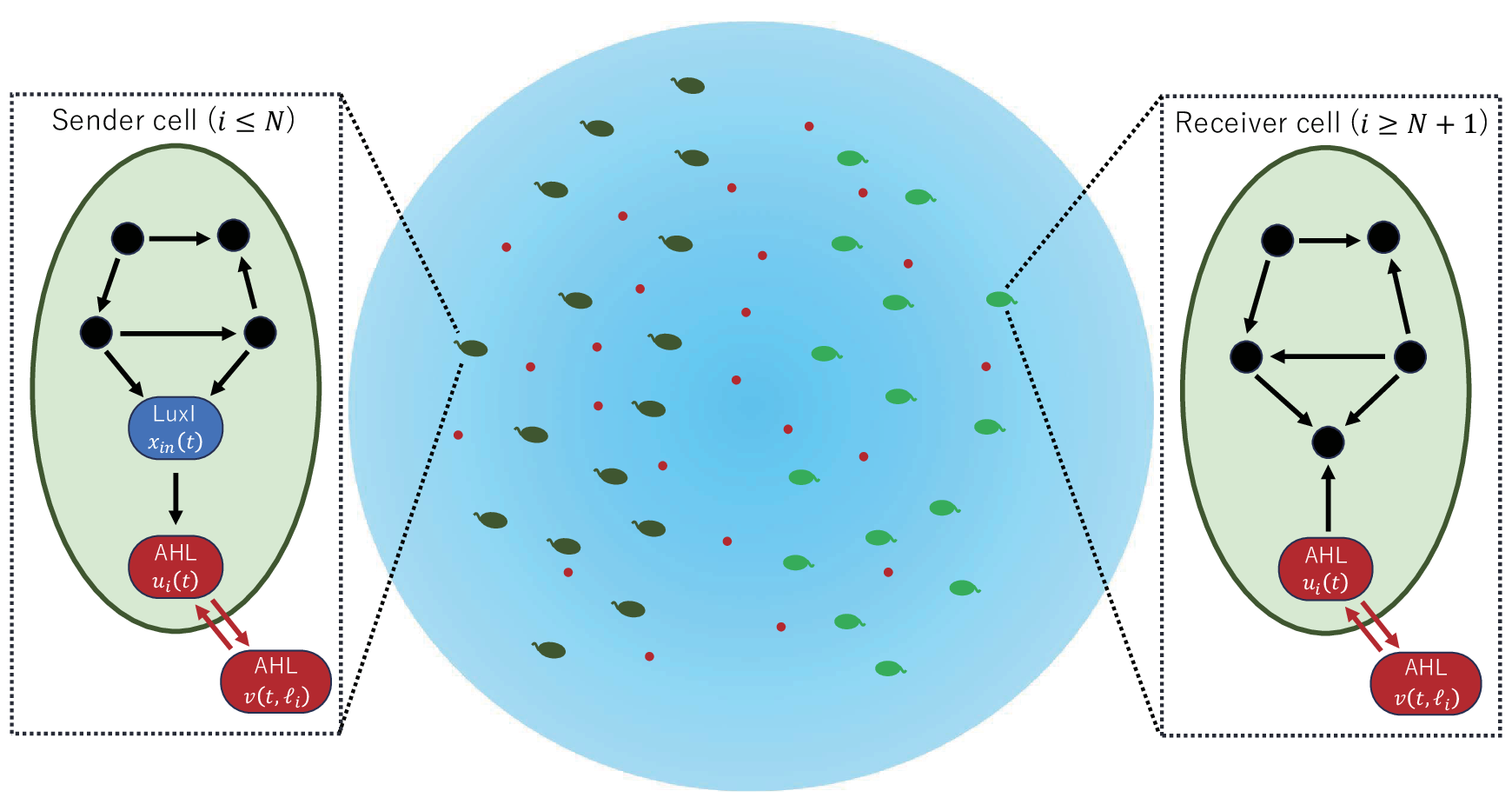}
    \caption{Schematic view of multicellular communication systems in hydrogel beads.}
    \label{fig:overview}
\end{figure}

We consider a multicellular communication system comprising a population of microscopic agents—such as engineered cells or synthetic protocells distributed within an aqueous medium (Fig. \ref{fig:overview}). The population is heterogeneous, consisting of \emph{sender cells} and \emph{receiver cells}. Sender cells secrete diffusible signal molecules (e.g., Acyl-homoserine lactone; AHL) synthesized by specific enzymes (e.g., LuxI). Receiver cells do not produce the signal but possess regulatory modules that sense the local concentration of the molecule and trigger intracellular responses.

Inside each agent, biochemical reactions occur in a well-mixed intracellular volume. The communication between these agents is mediated exclusively by the diffusive extracellular field, which transports molecules from senders to receivers. This architecture mimics natural quorum sensing but is engineered for programmable spatial behaviors.

Geometrically, the agents are embedded within a bounded domain (e.g., a hydrogel bead). The outer boundary of this domain acts as an absorbing interface, modeling the loss of signal molecules to the vast external environment. The agents are spatially separated such that direct physical contact does not occur, ensuring that all interactions are diffusion-mediated.

\subsection{Mathematical model}

To derive a tractable control-oriented model from the physical setup described above, we introduce the following geometric assumption.

\medskip
\begin{assumption}
\label{assu:model}
The physical size of each cell is negligible compared to the characteristic length scale of the outer domain.
\end{assumption}
\medskip

Under Assumption \ref{assu:model}, the volumetric flux from cells to the extracellular domain is approximated as a point source.

Let $i$ denote the cell index, where $i = 0, 1, \ldots, N$ correspond to sender cells, and $i = N+1, \ldots, N+M$ correspond to receiver cells. 
The overall system is modeled as a coupled PDE--ODE network. Sender and receiver cells have different intracellular dimensions, and we write
\[
\bm{x}_i(t)\in
\begin{cases}
\mathbb{R}^n, & i \le N \quad (\text{sender cell}),\\[4pt]
\mathbb{R}^m, & i \ge N+1 \quad (\text{receiver cell}),
\end{cases}
\]
represents the concentrations of molecular species in the $i$-th cell. We denote the $j$-th component of $\bm{x}_i(t)$ by $x_{i,j}(t)$, which represents the concentration of the $j$-th molecular species. The dynamics of the intracellular molecules is
\begin{equation}
\label{eq:xdynamics}
    \frac{d\bm x_i(t)}{dt} = \bm f(\bm x_i, u_i),
\end{equation}
where $u_i(t)$ is the concentration of AHL signal molecule in $i$-th cell. The local reaction term $\bm{f}(\cdot)$ is given by
\begin{equation}
\bm{f}(\bm{x}_i,u_i) = 
    \begin{cases}
        \bm f_s(\bm x_i, u_i) & i \le N,\\[4pt]
        \bm f_r(\bm{x}_i, u_i) & i \ge N+1,
    \end{cases}\label{eq:f}
\end{equation}
where $\bm{f_s}:\mathbb{R}^n \to \mathbb{R}^n$ describes the intracellular reaction dynamics in a sender cell, and $\bm{f_r}:\mathbb{R}^m \to \mathbb{R}^m$ describes the corresponding dynamics in a receiver cell.

For each cell $i\in\{1,\ldots,N+M\}$, we define the output variable $y_i(t)$, which represents the intracellular concentration of the AHL synthase LuxI. Since only sender cells produce LuxI, while receiver cells do not, we use the following unified definition:
\begin{equation}
    y_i(t) :=
    \begin{cases}
        \bm{c} \bm{x}_i(t), & i \le N,\\[6pt]
        0,          & i \ge N+1.
    \end{cases}
    \label{eq:yi_output}
\end{equation}
Here, $\bm{c}\in\mathbb{R}^{1\times n}$ is a constant row vector whose $n$-th entry is equal to $1$ and all other entries are zero, so that $\bm{c} \bm{x}_i(t)$ extracts the concentration of LuxI from the sender state vector $\bm{x}_i(t)$. 

\smallskip
\par
Next, consider the dynamics of signal molecules. 
Let $\bm{r}=[r_1, r_2, r_3]$ denote the spatial coordinate in three-dimensional spherical space $\Omega=\{\bm{r}\in\mathbb{R}^3:\|\bm{r}\|<L\}$ with the radius $L$, and
$v(t,\bm r)$ denote the concentration of the signal molecule at the position $\bm{r}$. The field evolves according to
\begin{equation}
\label{eq:diffusion}
\frac{\partial v(t,\bm r)}{\partial t}
= D\nabla^2 v(t,\bm r)
+ \sum_{i=1}^{N+M} \delta(\bm r - \bm \ell_i)\,
V\alpha \big(u_i(t) - v(t,\bm \ell_i)\big),
\end{equation}
where $D$ is the diffusion coefficient, $\delta(\cdot)$ denotes the Dirac delta function, and the constant \(\alpha\) represents the exchange rate between the intracellular and extracellular environments. The parameter $V$ is the cell volume $V=4\pi R^3/3$ with the cell radius $R$. Each Dirac delta function $\delta(\cdot)$ models a cell as a point-like source or sink, through which molecules are produced or absorbed according to the reaction rate $\alpha\bigl(u_i(t)-v(t,\bm{\ell}_i)\bigr)$. Thus, a cell with intracellular state $u_i(t)$ injects molecules into the environment when $u_i(t)>v(t,\bm{\ell}_i)$ and removes molecules when $u_i(t)<v(t,\bm{\ell}_i)$. In this way, Eq.~\eqref{eq:diffusion} captures the fundamental mechanism of diffusion-mediated communication: cells modify the extracellular concentration locally, and the resulting field propagates this information to other cells through diffusion. The dynamics of the intracellular concentration of the signal molecule $u_i(t)$ in the sender and the receiver cells is
\begin{equation}
    \frac{du_i(t)}{dt} = a_u y_i(t) - \gamma_u u_i(t) + \alpha(v(t,\bm{\ell}_i) - u_i(t)),\label{eq:IntracellularSignal}
\end{equation}
where \(\bm{\ell}_i\) is the spatial position of $i$-th cell.

Since the signal molecules that leave the microbead are rapidly dispersed by the surrounding flow, the Dirichlet boundary condition is imposed at the domain boundary as
\begin{equation}
v(t,\bm r)\big|_{\partial\Omega}=0.\label{eq:BoundaryCondition}
\end{equation}

While the ODE–PDE coupling model \eqref{eq:xdynamics}--\eqref{eq:BoundaryCondition} captures the spatially distributed diffusion dynamics and intracellular reaction processes with high fidelity, it poses several challenges for both analysis and computation. First, the infinite-dimensional nature of the diffusion PDE prevents direct application of standard tools from nonlinear systems theory, making stability, controllability, and input–output analysis difficult to carry out in closed form. Second, the coupling between the PDE field and a large number of cellular ODEs typically yields a stiff, high-dimensional dynamical system whose numerical simulation requires fine spatial discretization and small time steps, resulting in substantial computational cost. These difficulties motivate the development of reduced-order representations that retain the essential features of diffusion-mediated communication while enabling tractable analysis and efficient computation.

\section{Model reduction via singular perturbation}
\label{sec:singular}

To address the analytical and computational difficulties, we exploit the intrinsic time-scale separation between fast diffusion dynamics and slower intracellular reactions. We first show that the diffusion PDE converges exponentially to its quasi-steady spatial profile for any fixed intracellular state, ensuring that the field dynamics can be treated as a fast dynamics. Building on this property, we apply singular perturbation theory to rigorously reduce the coupled ODE–PDE system to a finite-dimensional network model with static interconnection. 

\subsection{Time-scale separation}
To formalize the hierarchy of dynamics within the system, we introduce characteristic time constants. The relaxation time of the diffusion equation \eqref{eq:diffusion} is governed by the domain size $L$ and the diffusion coefficient $D$. Specifically, associating the decay rate with the first Dirichlet eigenvalue of the spherical domain, we define the diffusion time constant $\tau_v$. Similarly, the time constant for the intracellular accumulation of the signal molecule \eqref{eq:IntracellularSignal} is determined by the membrane exchange rate $\alpha$, denoted as $\tau_u$. Thus, we define:
\begin{align}
    \tau_{v}
    := \frac{L^{2}}{\pi^2D}, \,\,\,\,& \,\,\,\, \tau_u := \frac{1}{\alpha}.
\end{align}
In the same way, we define $\tau_x$ as the time constant for the intracellular dynamics \eqref{eq:xdynamics}.

Based on the physical nature of biochemical communication, we impose the following assumption regarding the separation of these time scales.

\smallskip
\par
\begin{assumption}
    Consider the multicellular communication system \eqref{eq:xdynamics}--\eqref{eq:BoundaryCondition}. We assume the dynamics of the signal molecule relaxes much faster than the dynamics of the intracellular molecules:
    \label{eq:epsilon}
    \begin{align}
    \varepsilon_v \ll 1,\,\,\,\, & \,\,\,\, \varepsilon_u \ll 1,
    \end{align}
    where $\varepsilon_v = \tau_v/\tau_x$ and $\varepsilon_u = \tau_u/\tau_x$.
\end{assumption}

\smallskip
\par
This assumption is quantitatively justified under typical experimental conditions. For AHL molecules diffusing within a hydrogel of radius $L\simeq 300~\si{\micro m}$, the diffusion coefficient is typically $D\simeq(1\text{--}3)\!\times\!10^{4}~\si{\micro m^2/min}$
in aqueous or agarose environments \cite{Stewart2003}. Hence, the diffusion time scale is
\[
\tau_v = \frac{L^2}{\pi^2 D}
\;\approx\;
\frac{(300)^2}{\pi^2\times10^{4}}
\approx 1~\mathrm{min}.
\]
Regarding membrane transport, the permeability of AHL is reported as $\alpha\approx0.1\text{--}100\,\mathrm{min}$ \cite{Li2021}, which corresponds to a time constant $\tau_u \approx 0.01\text{--}10\,\mathrm{min}$.

In contrast, intracellular protein degradation typically occur over tens of minutes to hours, corresponding to $\tau_x\simeq10\text{--}100~\mathrm{min}$
\cite{Basu2005}. This significant gap validates Assumption \ref{eq:epsilon}, justifying the treatment of the diffusion field $v(t,\bm r)$ and intracellular signal $u_i(t)$ as fast dynamics.

Finally, to exploit this separation explicitly, we rescale time by $\hat{t} = t/\tau_x$ and space by $\hat{\bm{r}}=\bm{r}/L$. Under this scaling, the coupled system \eqref{eq:xdynamics}--\eqref{eq:IntracellularSignal} transforms into the standard singular perturbation form:
\begin{align}
    \label{eq:sf-pde}
    \varepsilon_v\,\frac{\partial v(\hat{t},\hat{\bm{r}})}{\partial \hat{t}} =&\, \nabla^{2} v(\hat{t},\hat{\bm{r}})\\
    &+ \frac{V\alpha}{LD} \sum_{i=1}^{N+M} \delta(\bm{\hat{r}} - \bm{\hat{\ell}_i})\, \big(u_i(\hat{t}) - v(\hat{t},\bm{\hat{\ell}_i})\big),\nonumber\\ 
    \varepsilon_u\frac{du_i(\hat{t})}{d\hat{t}} =& \frac{a_u}{\alpha}y_i(\hat{t}) - \frac{\gamma_u}{\alpha} u_i(\hat{t}) + (v(\hat{t},\bm{\hat{r}}_i) - u_i(\hat{t})),\label{eq:sf-u}\\
    \label{eq:sf-odes}
    \frac{d\bm x_i(\hat{t})}{d\hat{t}} =& \tau_x\bm f(\bm x_i, u_i),
\end{align}
These equations make explicit the hierarchical structure of the system, where fast diffusion, fast membrane exchange, and slow intracellular regulation operate on distinct time scales.

\subsection{Exponential stability of the diffusion system}
\label{sec:exp-stab}

Before performing the model reduction, we rigorously verify that the fast dynamics converges to a unique equilibrium for any fixed state of the slow subsystem. This exponential stability of the boundary-layer dynamics is a fundamental prerequisite for applying singular perturbation theory.

We study the boundary-layer PDE obtained by freezing the intracellular states at their initial values $(\bm x_i(0), y_i(0))$ and introducing the fast time scale variable $\tilde{t}=t/\tau_v$:
\begin{equation}
\label{eq:fast_v}
\begin{aligned}
    \frac{\partial v(\tilde{t},\hat{\bm{r}})}{\partial \tilde{t}} =&\, \nabla^2 v(\tilde{t},\bm{\hat{r}})\\ 
    &+ \frac{V\alpha}{LD} \sum_{i=1}^{N+M} \delta(\bm{\hat{r}} - \bm{\hat{\ell}_i})\, \big(u_i(\tilde{t}) - v(\tilde{t},\bm{\hat{\ell}_i})\big),
\end{aligned}
\end{equation}
\begin{equation}
    \frac{du_i(\tilde{t})}{d\tilde{t}} = \frac{L^2 a_u}{D}y_i(0) - \frac{L^2\gamma_u}{D} u_i(\tilde{t}) + \frac{L^2 \alpha}{D}(v(\tilde{t},\bm{\hat{r}}_i) - u_i(\tilde{t})). \label{eq:fast_u}
\end{equation}
Equation \eqref{eq:fast_u} exhibits exponential convergence since its linear dynamics have a negative decay rate. In the following proposition, we show that equation \eqref{eq:fast_v} likewise exhibits exponential convergence.

\medskip
\begin{proposition}[Exponential stability]
Consider the diffusion system (\ref{eq:fast_v}), and assume there exists a steady state $v^\ast \in H_0^1(\Omega)$, where $H_0^1(\Omega)$ is the Sobolev space with zero boundary values. Then for all $\tilde{t} \geq 0$, the following exponential decay holds:
\begin{equation}
\|v(\tilde{t},\bm{\hat{r}}) - v^*(\bm{\hat{r}})\|_{L^2(\Omega)} \le e^{-c\,\tilde{t}}\,\|v(0,\bm{\hat{r}}) - v^*(\bm{\hat{r}})\|_{L^2(\Omega)},
\end{equation}
where the decay rate $c$ satisfies $c \geq \lambda_1$, 
where $\lambda_1$ is the first Dirichlet eigenvalue of $-\Delta$ on $\Omega$.
\end{proposition}

\medskip
\begin{proof}
Define the perturbation from the steady state as $w(\tilde{t}, \hat{\bm{r}}) := v(\tilde{t}, \hat{\bm{r}}) - v^*(\hat{\bm{r}})$. Since both $v$ and $v^*$ satisfy the boundary conditions, the dynamics of $w$ are governed by
\begin{equation}
    \partial_t w = \nabla^2 w - \frac{V\alpha}{LD} \sum_{i=1}^{N+M} \delta(\hat{\bm{r}} - \hat{\bm{\ell}}_i) w, \quad w|_{\partial\Omega} = 0.
\end{equation}
Taking the $L^2$ inner product with $w$ and applying Green's first identity (integration by parts) yields the energy balance equation:
\begin{equation} \label{eq:energy_balance}
    \frac{1}{2}\frac{d}{dt}\|w(\tilde{t})\|_{L^2}^2 + a(w, w) = 0,
\end{equation}
where the bilinear form $a(\cdot, \cdot)$ is defined as
\begin{equation}
    a(w, w) := \int_\Omega |\nabla w|^2\,d\hat{\bm{r}} + \sum_{i=1}^{N+M}\frac{V\alpha}{LD}|w(\tilde{t}, \hat{\bm{\ell}}_i)|^2.
\end{equation}
To analyze the decay rate, we define the coercivity constant $c$ via the Rayleigh quotient:
\begin{equation} \label{eq:decay_c}
    c := \inf_{\varphi \in H_0^1(\Omega)\setminus\{0\}} \frac{a(\varphi, \varphi)}{\|\varphi\|_{L^2}^2}.
\end{equation}
Using the variational characterization of the first Dirichlet eigenvalue $\lambda_1$ of the Laplacian \cite{strauss2008partial}, we have the inequality $\|\nabla \varphi\|_{L^2}^2 \ge \lambda_1 \|\varphi\|_{L^2}^2$. Since the second term in $a(w, w)$ is non-negative, it follows that
\begin{equation}
    \frac{a(\varphi, \varphi)}{\|\varphi\|_{L^2}^2} 
    = \frac{\|\nabla \varphi\|_{L^2}^2}{\|\varphi\|_{L^2}^2} + \frac{\sum (V\alpha/LD)|\varphi(\hat{\bm{\ell}}_i)|^2}{\|\varphi\|_{L^2}^2} 
    \ge \lambda_1.
\end{equation}
Taking the infimum yields $c \ge \lambda_1 > 0$.
Finally, from \eqref{eq:decay_c}, we have $a(w, w) \ge c \|w\|_{L^2}^2$. Substituting this into \eqref{eq:energy_balance} gives the differential inequality:
\begin{equation}
    \frac{d}{dt}\|w\|_{L^2}^2 \le -2c \|w\|_{L^2}^2.
\end{equation}
Applying Grönwall's inequality results in $\|w(\tilde{t})\|_{L^2}^2 \le e^{-2c\tilde{t}}\|w(0)\|_{L^2}^2$, which proves the exponential stability.
\end{proof}

\subsection{Model reduction via singular perturbation}

Building on Proposition~1, we have established that the diffusion system~\eqref{eq:diffusion} constitutes a fast dynamics that converges exponentially to its quasi-steady state for any fixed intracellular variables. This validates the required time-scale separation for applying singular perturbation theory. In the 
following theorem, we show that the full ODE--PDE coupled system can therefore be rigorously reduced to a finite-dimensional network system with a static interconnection map, yielding an analytically tractable representation of the multicellular communication dynamics.

For later use, define the input and output vectors
\begin{align*}
    \bm{U}(t) &:= [u_1(t),\ldots,u_{N+M}(t)]^{\top},\\
    \bm{Y}(t) &:= [y_1(t),\ldots,y_{N+M}(t)]^{\top}.
\end{align*}

\medskip
\begin{theorem}
    Consider the multicellular communication system \eqref{eq:xdynamics}--\eqref{eq:BoundaryCondition} under Assumption 1 and 2, and suppose that the system satisfies Proposition 1. Then, the system is reduced to the intracellular system \eqref{eq:xdynamics}--\eqref{eq:yi_output} with
    \begin{equation}
        \bm{U}(t) = \mathcal{G} \bm{Y}(t)
        \label{eq:reduced}
    \end{equation}
    where $\mathcal{G}$ is the \emph{communication gain matrix} given by
    \begin{equation}
        \mathcal{G} = \left(I - \frac{V\alpha^2}{\alpha + \gamma_u}G\left(I + V\alpha G \right)^{-1} \right)^{-1}\frac{a_u}{\alpha + \gamma_u}.\label{eq:G}
    \end{equation}
    The matrix $G \in \mathbb{R}^{(N+M)\times(N+M)}$ is defined by
    \begin{equation}
    \label{eq:G_matrix}
    \begin{aligned}
        G &= \begin{bmatrix} g(\bm{r_1},\bm{r_1})&\cdots&g(\bm{r_1},\bm{r_{N+M}})\\
        \vdots&\ddots&\vdots\\
        g(\bm{r_{N+M}},\bm{r_1})&\cdots&g(\bm{r_{N+M}},\bm{r_{N+M}})
        \end{bmatrix},
    \end{aligned}
    \end{equation}
    with 
    \begin{equation}
        \label{eq:g_hat}
        \begin{aligned}
            g(\bm{\ell}_j,\bm{\ell}_i) =
            \begin{cases}
            \frac{1}{4\pi D}\!\left(\frac{1}{\|\bm{\ell}_j-\bm{\ell}_i\|} - \frac{L}{\|\bm{\ell}_i\|}\, \frac{1}{\|\bm{\ell}_j - \bm{\ell}_i^*\|} \right), & j\neq i,\\[4pt]
            \displaystyle \frac{3}{8\pi DR} - \frac{L}{4\pi(L^2 - \|\bm{\ell}_i\|^2)}, & j=i.
            \end{cases}
        \end{aligned}
    \end{equation}
    and
    \begin{equation*}
        \bm{\ell}_i^* = \frac{L^2}{\|\bm{\ell}_i\|^2}\,\bm{\ell}_i.
    \end{equation*}
\end{theorem}

\medskip
\begin{proof}
In a limit of $\varepsilon_v \rightarrow 0$, the extracellular diffusion~\eqref{eq:sf-pde} becomes the elliptic equation
\begin{equation}
D\nabla^2 v(t,\bm r)
+ \sum_{i=1}^{N+M} \delta(\bm r - \bm \ell_i)\,
V\alpha \big(u_i(t) - v(t,\bm \ell_i)\big) = 0.
\label{eq:elliptic}
\end{equation}

\par
\smallskip
Let $g(\bm r,\bm \ell_i)$ denote the Green’s function satisfying
\begin{equation*}
    D\nabla^2 g(\bm r,\bm \ell_i) = -\delta(\bm r-\bm \ell_i),
\qquad g|_{\partial\Omega}=0.
\end{equation*}
As shown in \cite{jackson_classical_1999}, the Green's function $g(\bm r,\bm \ell_i)$ is obtained as
\begin{equation}
    g(\bm r,\bm \ell_i) = \frac{1}{4\pi D}\!\left(\frac{1}{\|\bm{r}-\bm{\ell}_i\|} - \frac{L}{\|\bm{\ell}_i\|}\,    \frac{1}{\|\bm{r} - \bm{\ell}_i^*\|} \right).
    \label{eq:green_g}
\end{equation}
For $\bm r=\bm \ell_j \, (j\neq i)$, we have
\begin{equation*}
    g(\bm \ell_j,\bm \ell_i)=\frac{1}{4\pi D}\!\left(\frac{1}{\|\bm \ell_j-\bm \ell_i\|} - \frac{L}{\|\bm \ell_i\|}\frac{1}{\|\bm \ell_j-\bm \ell_i^*\|}
\right).
\end{equation*}
Since the first term in \eqref{eq:green_g} is singular at $\bm r=\bm \ell_j$, We regularize the self-interaction by averaging $g(\cdot,\cdot)$ over the small cell region:
\begin{equation*}
\begin{aligned}
    g(\bm{\ell_j},\bm{\ell_j}) &\approx \left[\frac{3}{4\pi R^3}\int_{B_R(\bm \ell_j)}\!\! g(\bm r,\bm{\ell_j})\,d\bm r\right]_{\bm{r}=\bm{\ell_j}}\\
    &= \frac{3}{8\pi DR} - \frac{L}{4\pi D(L^2 - \|\bm{\ell}_j\|^2)},
\end{aligned}
\end{equation*}
where \( B_R(\bm{\ell_j}) \) denotes the small cell region of radius \( R \) centered at \( \bm{\ell_j} \), i.e.,
\[
B_R(\bm{\ell_j}) = \{\, r \in \mathbb{R}^3 \mid \| \bm{r} - \bm{\ell_j} \| \le R \,\}.
\]
Combining both cases yields Eq. \eqref{eq:g_hat}.

Now, utilizing the linearity of the elliptic equation~\eqref{eq:elliptic}, the general solution $v(t, \bm{r})$ can be expressed as a superposition of the contributions from all point sources via the Green's function:
\begin{equation}
    v(t,\bm{\ell_i}) = V\alpha \sum_{j=1}^{N+M} g(\bm{\ell_i},\bm{\ell_j}) \big(u_i(t) - v(t,\bm \ell_j)\big).
    \label{eq:OpenLoopScalar}
\end{equation}
Let $\bm{\nu}(t)\in\mathbb{R}^{N+M}$ denote the vector whose $i$-th entry is $v(t,\bm{\ell_i})$. Then, the vector form of Eq. \eqref{eq:OpenLoopScalar} is
\begin{equation*}
\begin{aligned}
    \bm{\nu}(t) = V\alpha G (\bm{U}(t) - \bm{\nu}(t)).
\end{aligned}
\end{equation*}
This leads to 
\begin{equation}
    \bm{\nu}(t) = V\alpha G(I+ V\alpha G)^{-1} \bm{U}(t).
    \label{eq:v_g}
\end{equation}

In the limit of $\varepsilon_u \rightarrow 0$, the left hand side of the intracellular dynamics of the signal molecule \eqref{eq:sf-u} is zero as
\begin{equation*}
    a_u y_i(t) - \gamma_u u_i(t) + \alpha(v(t,\bm{\ell}_i) - u_i(t)) = 0,
\end{equation*}
which leads to 
\begin{equation}
    \bm{U}(t) = \frac{a_u}{\alpha + \gamma_u}\bm{Y}(t) + \frac{\alpha}{\alpha + \gamma_u}\bm{\nu}(t)
    \label{eq:U_open}
\end{equation}
in a vector form. Substituting Eq. \eqref{eq:v_g} into Eq. \eqref{eq:U_open}, we obtain Eq. \eqref{eq:reduced}.

\end{proof}

From a control perspective, Theorem~1 replaces the coupled PDE--ODE description \eqref{eq:xdynamics}--\eqref{eq:BoundaryCondition} with the finite-dimensional network system \eqref{eq:xdynamics}--\eqref{eq:yi_output} through the static interconnection $U(t)=\mathcal{G}Y(t)$. The matrix $\mathcal{G}$ provides a closed-form representation of the combined effects of diffusion, membrane transport, and degradation, embedding the spatio--temporal coupling of the diffusion field into a time-invariant communication gain. This enables systematic analysis of how geometry and physical parameters shape the effective interaction
topology, while allowing the multicellular network to be viewed as a standard multi-agent system. As a consequence, classical tools such as Lyapunov and ISS methods \cite{Khalil2002,Sontag2008}, small-gain and passivity criteria \cite{desoer1975feedback}, and graph-theoretic approaches to synchronization and pattern formation \cite{Olfati2004,Mesbahi2010graph} become applicable without handling the infinite-dimensional PDE dynamics directly.

Moreover, the interconnection $U(t)=\mathcal{G}Y(t)$ significantly reduces computational complexity, as simulations and controller design can be performed on a system of size proportional to the number of cells rather than on a high-dimensional discretization of the diffusion equation. Finally, since the diffusion
subsystem is exponentially stable, classical singular perturbation results guarantee that the trajectories of the reduced model approximate those of the full PDE--ODE system with an error of order $O(\varepsilon)$ \cite{Khalil2002}, thus justifying the use of the reduced network model for the subsequent analysis.

\section{Numerical Example}
\label{sec:numerical}

In the following numerical example, we consider a multicellular
communication system composed of sender and receiver cells as shown in Fig. \ref{fig:error}A. In each sender cell, the synthase LuxI is constitutively expressed and produces the signal molecule AHL. The AHL molecules diffuse across the cell membrane and propagate through the extracellular domain. Each receiver cell contains a genetic toggle switch composed of the transcriptional repressors LacI and TetR. The incoming AHL activates the expression of TetR, thereby biasing the toggle switch. As a result, receiver cells exposed to sufficiently high AHL levels settle in a high-TetR/low-LacI state, whereas those receiving little AHL remain in a low-TetR/high-LacI state. This enables the receiver cells to infer the presence or absence of the sender signal through their intracellular toggle-switch state.

In the sender cells ($0\leq i\leq N$), the synthase LuxI is produced as:
\begin{equation}
    \dot{x}_{i1}(t) = a_s - \gamma_s x_{i1}(t), \label{eq:ex_sender}
\end{equation}
where $a_s$ and $\gamma_s$ are the production rate and the degradation rate of $x_{i1}$, respectively. 
The receiver cells ($N+1\leq i \leq N+M$) have two molecular species: LacI ($x_{i1}$) and TetR ($x_{i2}$) whose dynamics is
\begin{equation}
    \begin{aligned}
        \dot{x}_{i1}(t) &= a_{r1} \frac{K_2^2}{K_2^2 + x_{i2}^2(t)} - \gamma_{r1} x_{i1}(t), \\
        \dot{x}_{i2}(t) &= a_{r2}\!\left(\frac{u_{i}^2(t)}{K_u^2 + u_{i}^2(t)} + \frac{K_1^2}{K_1^2 + x_{i1}^2(t)}\right)
                    - \gamma_{r2} x_{i2}(t),
    \end{aligned}\label{eq:ex_receive}
\end{equation}
where $a_{rj}$ and $\gamma_{rj}$ are the production rate and the degradation rate, respectively. The parameter $K_j$ is the dissociation constant. 
The dynamics of the extracellular and intracellular AHL signal molecule is modeled by Eqs.~\eqref{eq:diffusion}--\eqref{eq:IntracellularSignal}.

In the numerical simulations, we use the following parameter values. The intracellular molecular concentrations are measured in nM. The production rates are set to $a_1 = 5.0\,\mathrm{nM/min}$, $a_2 = 5.0\,\mathrm{nM/min}$, and $a_3 = 2.5\,\mathrm{nM/min}$, while the degradation rates are $g_1 = g_2 = g_3 = g_A = 0.01\,\mathrm{min^{-1}}$. The AHL-related production and degradation use $a_A = 2.0\,\mathrm{nM/min}$ and $g_A = 0.01\,\mathrm{min^{-1}}$. The Hill-function dissociation constants are dimensionless and chosen as $K_2 = 50.0\,\mathrm{nM}$, $K_3 = 50.0\,\mathrm{nM}$, and $K_4 = 10.0\,\mathrm{nM}$. The membrane exchange parameter is set to $\alpha = 1.0\,\mathrm{min^{-1}}$. The cell radius is $R = 1.5\,\si{\micro m}$, and the diffusion coefficient of the signal molecule is $D = 2.0\times10^{4}\,\si{\micro m^2/min}$.

\subsection{Error Evaluation of the Reduced Model}

\begin{figure}
    \centering
    \includegraphics[width=0.99\linewidth]{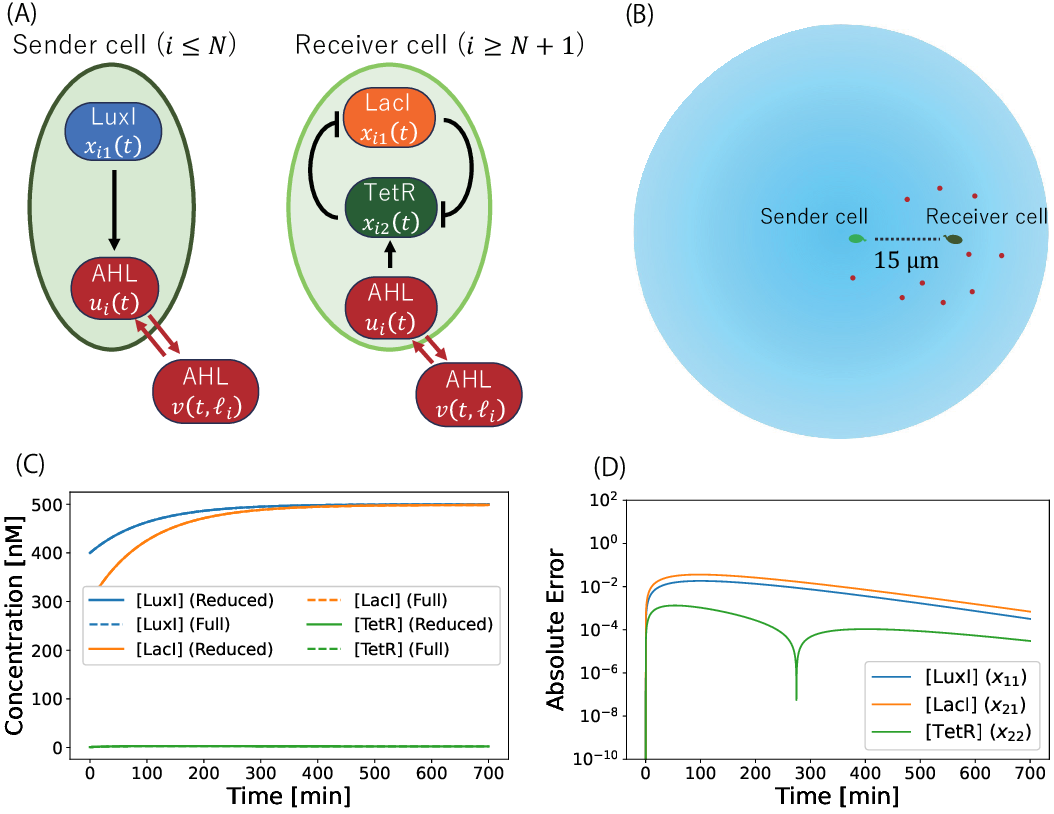}
    \caption{(A) Genetic circuits in sender cells and receiver cells. (B) Spatial arrangement of sender cells and receiver cells. (C) Dynamics of the intracellular molecular concentration for the full model and the reduced model. (D) Dynamics of the absolute error.}
    \label{fig:error}
\end{figure}

We first validate the reduced model in Theorem~1 by numerically comparing it with the full ODE--PDE model. To reduce the computational cost of the PDE simulation, we set the half-domain size to $L = 20\,\si{\micro m}$ (note that this is smaller than realistic hydrogel beads to ensure computational tractability). We consider the simplest configuration with one sender cell and one receiver cell ($N = 1$, $M = 1$). The sender cell is placed at the origin, and the receiver cell is located at $[15,\,0,\,0]\,\si{\micro m}$ as shown in Fig. \ref{fig:error}B. We set the initial values as $x_{11}(0)=400\,\mathrm{nM}$, $x_{21}(0)=300\,\mathrm{nM}$, $x_{22}(0)=1\,\mathrm{nM}$. Note that the high initial value of [LacI] ($x_{21}$) represents the OFF state of the receiver cell.

For the chosen parameter set, the singular perturbation parameters are calculated as $\varepsilon_u=0.01$ and $\varepsilon_v = 2.2\times 10^{-5}$. Since these values satisfy the time-scale separation condition, we apply Theorem~1 to obtain the reduced model. Figure~\ref{fig:error}A illustrates the temporal evolution of the intracellular molecular concentrations for both the reduced model and the full ODE--PDE model. Solid curves correspond to the reduced model, while dashed curves indicate the trajectories obtained from the full model. As shown in the figure, the two trajectories almost overlap for all molecular species, indicating that the reduced model accurately reproduces the dynamics of the full system. Figure~\ref{fig:error}B shows the absolute errors between the two models over time. The errors remain small ($<10^{-1}$) throughout the simulation horizon. To further examine the dependence of the approximation accuracy on the time-scale parameters, Table~1 summarizes the maximum errors obtained for several choices of $\varepsilon_u$ and $\varepsilon_v$ by changing the degradation rate $\gamma_j\,(j=s,r1,r2)$. The observed errors vary approximately linearly with respect to $\varepsilon_u$ and $\varepsilon_v$, which is consistent with the theoretical error bound predicted by singular perturbation theory.

\begin{table}[t]
\centering
\caption{Maximum absolute error for different values of $\varepsilon$.}
\label{tab:epsilon_error}
\begin{tabular}{c|ccc}
\hline
$\varepsilon_u$, $\varepsilon_v$ & $0.1$ & $0.05$ & $0.01$ \\
\hline
\makecell{\rule{0pt}{3ex}Absolute errors \\ of $[x_{21},\, x_{22}]^\top$} & \makecell{$\begin{bmatrix}4.6\times10^{-1}\\ 3.3\times10^{-2}\end{bmatrix}$} & \makecell{$\begin{bmatrix}1.8\times10^{-1}\\ 1.1\times10^{-2}\end{bmatrix}$} & \makecell{$\begin{bmatrix}3.6\times10^{-2}\\ 1.3\times10^{-3}\end{bmatrix}$} \\
\hline
\end{tabular}
\end{table}

It is also noteworthy that, in this particular configuration, the signal emitted from the sender cell does not reach the receiver cell at sufficiently high levels. Consequently, the concentration [TetR] ($x_{i2}$) inside the receiver cell does not exhibit activation. This behavior is consistent across both models, further confirming that the reduced model captures the qualitative behavior of the full system.

\subsection{Analysis of Spatial Arrangement Dependence}

Next, we analyze the dependence of the multicellular system on the spatial arrangement of cells. The radius of the hydrogel bead (i.e., the spatial domain) is set to $L = 300\,\si{\micro m}$, which is large enough to accommodate multiple sender cells at different locations. We perform simulations to examine how the spatial configuration of $4000$ sender cells affect the response of a single receiver cell. The initial values are same as the previous simulation setting. 

\begin{figure}
    \centering
    \includegraphics[width=0.99\linewidth]{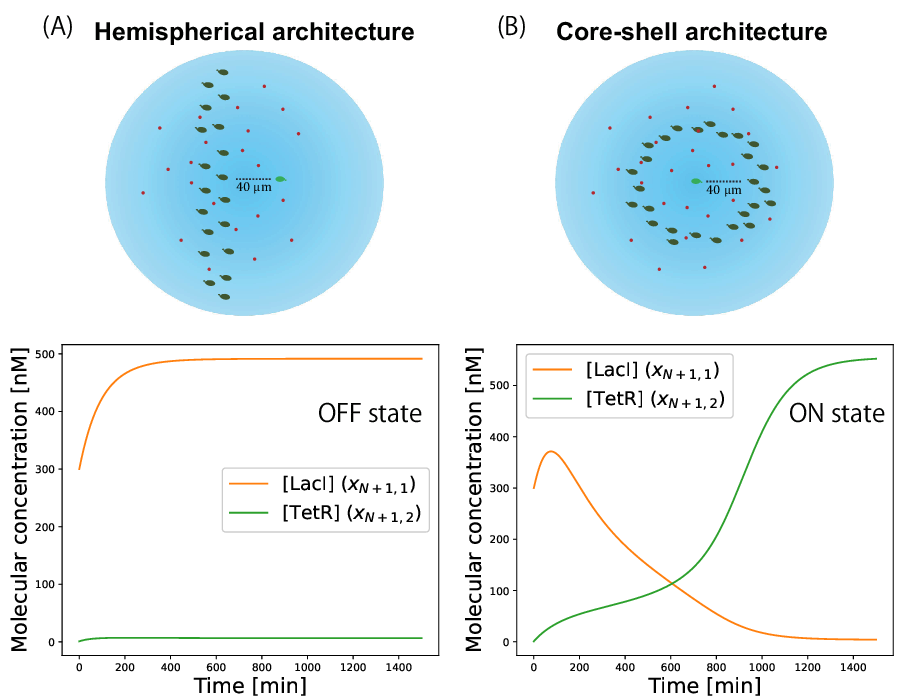}
    \caption{Dynamics of the molecular concentration in the receiver cell (A) when 4000 sender cells are uniformly distributed within a thin slab on the left hemisphere while the receiver cell is placed at $[40,0,0]\,\si{\micro m}$ on the right side of the domain, and (B) when 4000 sender cells are distributed approximately uniformly within a spherical shell of radius $40\,\si{\micro m}$ while the receiver cell is placed at the origin.}
    \label{fig:spatialanalysis}
\end{figure}

To examine how the spatial distribution of sender cells influences the receiver’s toggle-switch response, we compare two distinct spatial configurations within a spherical hydrogel bead. The first configuration (Fig. \ref{fig:spatialanalysis}A) mimics a hemispherically separated architecture \cite{yoshida2017comp}, where sender cells are distributed uniformly within a slab defined by $-20\,\si{\micro m} \le r_1 \le 0\,\si{\micro m}$ with the receiver cell placed remotely at $[40,0,0]\,\si{\micro m}$. The second configuration (Fig. \ref{fig:spatialanalysis}B) represents a core-shell encapsulation architecture \cite{sousa2023hierarchical,Jeong2023}, where the same number of sender cells form a spherical shell of radius $40\,\si{\micro m}$ surrounding the receiver cell located at the origin. Under this parameter set, we calculate the singular perturbation parameters as $\varepsilon_u=0.01$ and $\varepsilon_v = 5.0\times 10^{-3}$. These values sufficiently satisfy the time-scale separation, justifying the use of the reduced model. Consequently, the intracellular dynamics are simulated using the static interconnection defined by the communication gain matrix $\mathcal{G}\in\mathbb{R}^{4001\times 4001}$, computed via Eq.~\eqref{eq:G}.

Figure~\ref{fig:spatialanalysis} shows the receiver’s molecular trajectories for the different spatial arrangement of the cells. In Fig. \ref{fig:spatialanalysis}A, the AHL signal arriving at the receiver remains below the activation threshold, and the toggle switch stays in the OFF state. In contrast, in Fig. \ref{fig:spatialanalysis}B, the aggregated AHL influx becomes sufficiently large to trigger the switch, resulting in a transition to the ON state. This threshold-like behavior reflects how cooperative signaling among spatially distributed senders determines whether long-range molecular communication is successful. This result implies that a concentric core-shell arrangement enables significantly more efficient signal transmission from sender to receiver cells compared to a hemispherically separated configuration in hydrogel beads. This highlights spatial organization as a critical design parameter for overcoming diffusion limits in engineered multicellular systems.

This case study validates the scalability of our model reduction method. By replacing the stiff PDE-ODE coupling with a static interconnection map, the reduced model facilitates the simulation of thousands of agents. This capability is essential for the design and analysis of engineered multicellular systems where spatial density and heterogeneity play dominant roles.

\section{Conclusion}

In this paper, we have analyzed multicellular systems that communicate through diffusible signal molecules and are modeled by ODE–PDE coupling dynamics. Using singular perturbation theory, we have proposed a rigorous reduction method that transforms these infinite-dimensional systems into finite-dimensional network models with static interconnection. Numerical comparisons with the full ODE–PDE dynamics confirmed that the approximation error scales linearly with the perturbation parameters and remains small across biologically relevant regimes. We have further demonstrated that the reduced model faithfully captures changes in receiver responses arising from different spatial arrangements of sender cells.

The proposed framework provides a systematic and analytically tractable representation of diffusion-mediated intercellular communication, enabling the use of conventional tools from networked control and dynamical systems theory. By replacing complex spatial communication dynamics with low-dimensional models, our approach facilitates analysis and design of engineered multicellular systems, including synthetic microbial consortia and hydrogel-based communication platforms.


\addtolength{\textheight}{-12cm}


\appendix

\end{document}